# OSL dosimetric properties of cerium doped lutetium orthosilicates

A. Twardak[1], P. Bilski[1], Y. Zorenko[2,3], V. Gorbenko[3], O. Sidletskiy[4]

[1]Institute of Nuclear Physics, Polish Academy of Science, 31-342 Krakow, Poland

[2]Institute of Physics, Kazimierz Wielki University in Bydgoszcz, 85-090 Bydgoszcz, Poland

[3]Electronics Department, Ivan Franko National University of Lviv (LOM), 79017 Lviv, Ukraine

[4]Institute for Scintillation Materials National AS of Ukraine (ISMA), 61001 Kharkiv, Ukraine

**Corresponding author:**
Anna Twardak
e-mail: anna.twardak@ifj.edu.pl
phone: +48126628490

**Abstract**

This paper presents selected optically stimulated luminescence properties of $Lu_2SiO_5$:Ce single crystalline films grown using Liquid Phase Epitaxy technique. Comparison of continuous wave optically stimulated luminescence decay curves under blue and green light stimulation is shown. The dose response characteristic is found to be linear in the studied range from 100 μGy to 1 Gy. Analyses of the linearly modulated optically stimulated luminescence signal enabled establishing of the photoionization cross sections for blue light (470 nm). Bleachability and thermal stability of CW-OSL signal are discussed, as well as preliminary results of the fading study.

**Introduction**

Optically stimulated luminescence (OSL) is one of the widely used methods of passive dosimetry (Yukihara et. al, 2011). However, it is a relatively new method and a need for characterizing OSL properties of potential OSL materials still remains. In this work we have performed analysis of OSL properties of cerium doped lutetium orthosilicates $Lu_2SiO_5$:Ce (LSO:Ce). LSO:Ce is a well-known and commonly used scintillator material (Suzuki et al. 1992, Melcher et al. 1993). LSO:Ce is applied in many fields, for example in Positron Emission Tomography (PET). For commercial applications single crystals of LSO:Ce grown by Czochralski method are used. Recently several studies have been done on luminescent properties of single crystalline films prepared using Liquid Phase Epitaxy (Zorenko et al. 2012). As was suggested by Zorenko et. al (2011) luminescence characteristic of LSO:Ce single crystalline films (SCF) are determine by $Ce^{3+}$ and $Pb^{2+}$ based centres (Pb ions coming from PbO flux used during the growth). Moreover these properties of LSO:Ce samples are also influenced by $Pb^{2+}$-$Ce^{3+}$ energy transfer process. It was shown that LSO:Ce single crystals (SC) and single crystalline films possess significant OSL signal under blue light stimulation

(Twardak et al. 2013). In this work further investigation on OSL properties of LSO:Ce SCF are presented.

**Materials and methods**

Optically stimulated luminescence was measured using two different OSL readers: Risø DA-20 TL/OSL reader (Risø DTU, Denmark) and a portable OSL reader Helios-1 (Mandowski et al. 2010). Optically stimulation system in Risø reader is based on 28 blue LEDs with peak emission at 470 nm. The total power density of all LEDs is 80 mWcm$^{-2}$ at sample surface. In detection system band pass filter U-340 (Hoya) enables measurement in The UV range (300-400 nm). Using this reader we measure the optically stimulated luminescence in two basic modes: continuous wave OSL (CW-OSL) and linear modulation OSL (LM-OSL). Helios-1 enables optical stimulation in the range of 520 to 535 nm. It is equipped with 5 green LEDs with a total power of 25 W. The optically stimulated luminescence can be measured in blue and the UV range. In this case only CW-OSL curves were registered. The samples were irradiated using beta source Sr-90/Y-90 built-in DA-20 reader. To erase any residual OSL signal, before each measurement, the samples were annealed in a reader in 450 $^{o}$C. In all stages of all experiments investigated samples were kept in the darkness, at room temperature. Thermoluminescence glow curves were registered using DA-20 reader with the heating rate 5 $^{o}$C/s.

Single crystalline films of cerium doped lutetium orthosilicate were prepared using liquid phase epitaxy (LPE) in LOM, Ukraine [Zorenko et. al 2011]. They were grown from super-cooled melt-solution based on the PbO-B$_2$O$_3$ flux onto yttrium orthosilicate (YSO) substrates. Cerium dopants were added in form of CeO$_2$ activator oxide ions to the melt solution. Their concentrations were 5, 10 or 20 mol%. The thicknesses of single crystalline films were between 8.7 and 37 μm. The investigated samples were grown for scintillation studies, so that their properties were not optimized for optically stimulated luminescence application. Aluminium oxide doped with carbon (Al$_2$O$_3$:C) obtained from Landauer was used as a reference material.

After preliminary measurements the sample with the most promising OSL signal intensity was selected and subjected to further investigations. This sample is LSO doped with 5 mol% of CeO$_2$ with single crystalline film thickness equal to 21 μm. The sample mass was 88.3 mg.

**Results and discussion**

Figure 1 presents the comparison of CW-OSL decay curves of LSO:Ce and Al$_2$O$_3$:C registered under green light stimulation. The initial OSL intensity per unit mass of LSO:Ce sample is nearly four times higher than that of aluminium oxide. Nevertheless the CW-OSL signal integral of Al$_2$O$_3$:C is 60% higher than that of lutetium orthosilicate due to the faster decay of LSO:Ce CW-OSL decay curve.

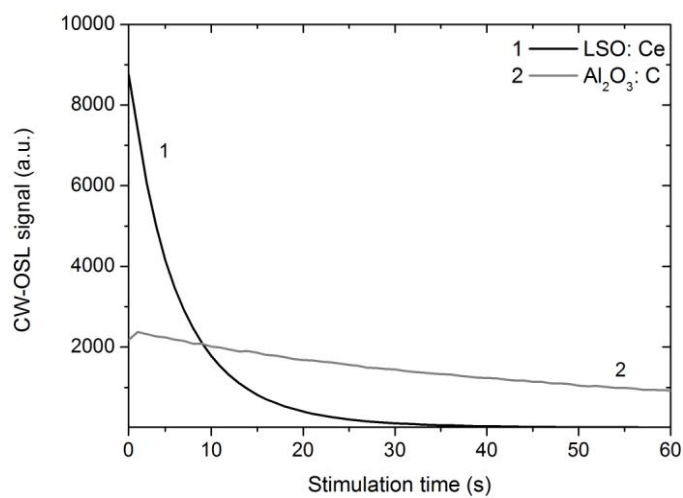

Fig. 1. CW-OSL decay curves of LSO:Ce and Al$_2$O$_3$:C under green light stimulation after irradiation with beta particles (210 mGy). Results were normalized to unit mass.

The intensity of CW-OSL signal of LSO:Ce is also higher than that of Al$_2$O$_3$:C under blue light stimulation (Figure 2). In this case the initial intensities differ two times, but signal integral of aluminium oxide is almost five times higher. It should be noted that detection in the UV range is not optimal for Al$_2$O$_3$:C, although it also does not match the emission spectra of LSO:Ce, which has a broad peak extending from about 350 nm to 500 nm (maximum near 430 nm). The CW-OSL decay curve under blue light stimulation is faster than that for stimulated with green light. It may be the result of both: the different photoionization cross section and different photon flux which depends on total power and stimulation light wavelength.

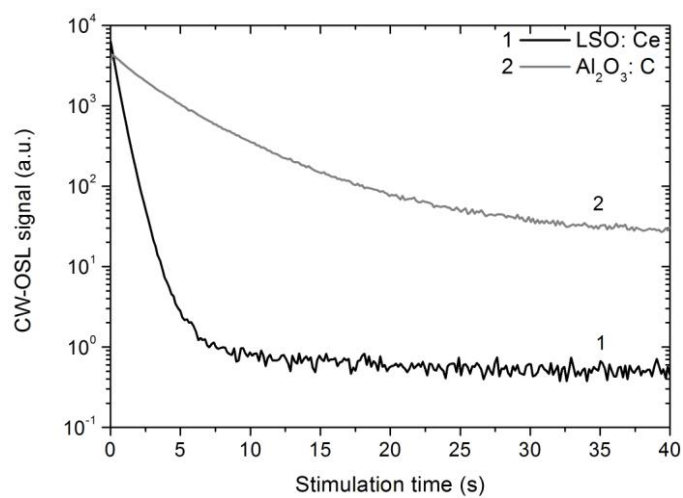

Fig. 2. CW-OSL decay curves of LSO:Ce and Al$_2$O$_3$:C previously exposed to beta particles (210 mGy) under blue light stimulation. Results were normalized to unit mass.

LM-OSL were measured using 100s stimulation time and 90% of LEDs power. The LM-OSL curves were fitted using add-in utility – Solver for Microsoft Excel (Afouxenidis et al. 2012) with three components using first order kinetic equation (Kitis and Pagonis 2008) (see Fig.3). The shape of single LM-OSL peak is characterised by:

$$I(t) = 1.6487 I_m \frac{t}{t_m} \exp\left(\frac{-t^2}{2t_m^2}\right) \qquad (1)$$

where $I_m$ – maximum peak intensity [a.u.], $t_m$ – time corresponding to peak maximum [s].

For each component the time of peak maximum ($t_{max}$) was established. Using this parameter the detrapping probability (b) can be calculated from the expression:

$$t_m = \sqrt{\frac{T}{b}} \qquad (2)$$

where T is total illumination time. Method of calculation of photoionization cross section (σ) was described by Choi (Choi et al. 2006). Photoionization cross section is related to detrapping probability and maximum stimulation photon flux (*J*):

$$b = J\sigma \qquad (3)$$

For stimulation wavelength 470 nm and power density 72 mWcm$^{-2}$ used in the experiment the calculated photon flux was $1.7 \cdot 10^{17}$ s$^{-1}$cm$^{-2}$. Using above parameters, the photoionization cross sections for all components were established. The results are presented in Table 1. As can be noticed, the greatest contribution to the whole curve gives the fast component.

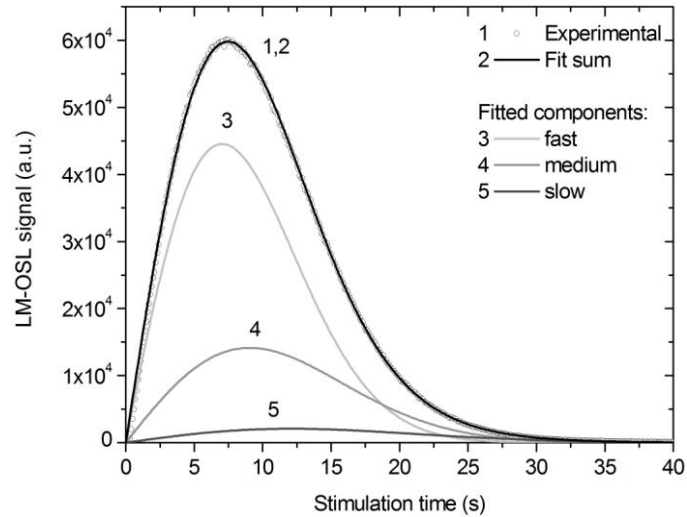

Fig. 3. LSO:Ce LM-OSL curves fitted with three component: fast, medium and slow. FOM% = 1.42

Table 1. LM-OSL peak maximum and corresponding photoionization cross section.

|  | Fast | Medium | Slow |
|---|---|---|---|
| $t_{max}$ (s) | 7.0 | 9.2 | 11.4 |
| σ (cm$^2$) | $1.19 \cdot 10^{-17}$ | $6.91 \cdot 10^{-18}$ | $4.55 \cdot 10^{-18}$ |
| Contribution (%) | 67 | 27 | 5 |

To study the dose response characteristic, the sample was irradiated with different doses of beta particles ranging from 100 μGy to 1 Gy. As can be seen in Figure 4 the CW-OSL signal is proportional to the absorbed radiation dose in the whole studied range. It should be noted that doses even below 1 mGy can be measured using LSO:Ce single crystalline film samples.

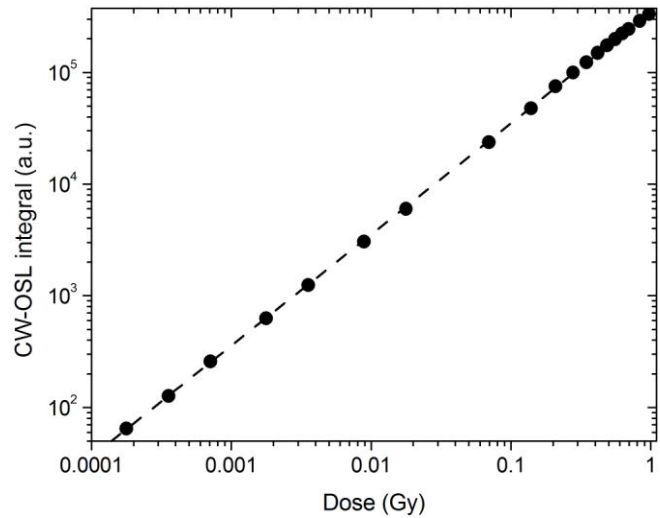

Fig. 4. Dose response characteristic after irradiation with beta particles.

The bleachability with blue light was tested. Figure 5 presents the residual CW-OSL signal integral after bleaching with blue light (470 nm) versus the bleaching time. A sample was annealed and then irradiated with 280 mGy of beta particles. Before each readout LSO:Ce was bleached at room temperature with 90% of total power of blue light stimulation source for the specified time in the range of 1 to 7200 s. The CW-OSL integral decreases rapidly even after one second of bleaching. Fast decrease of signal continuous for the next 10 seconds of bleaching and then starts to stabilize. Further bleaching results in a slower decrease of the CW-OSL decay curves integral and only after 5400 s reaches the background level (dashed line). This suggests that the slowest component of the OSL decay curves is hardly bleached and using blue light stimulation to erase the CW-OSL signal may be inefficient and may be achieved by thermal annealing.

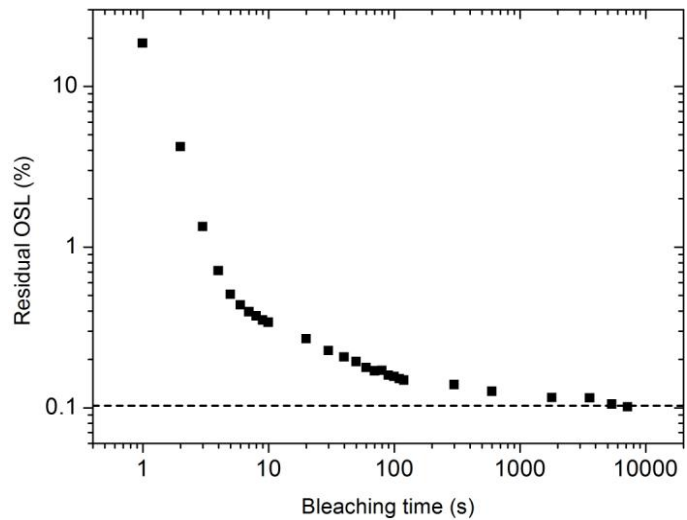

Fig. 5. Bleachability under blue light stimulation. OSL residual signal integral versus bleaching time. The background level is indicated with a dashed line. The measurements were conducted in room temperature.

In order to determine the thermal stability of the OSL signal the CW-OSL integral dependence on preheat temperature was measured. A sample previously annealed at 450 °C was irradiated with a dose equal to 280 mGy. In the next step preheat was done with linear heating with heating rate of

5°C/s up to temperatures from 80 °C to 300 °C with 10 °C step. Next the CW-OSL signal was registered for 40 s. Obtained results are presented in Figure 6. The CW-OSL integral started to decrease slowly after preheating to 100 °C. Then the decrease accelerates and after 160 °C the residual OSL signal is about 10% of the beginning signal. After this temperature CW-OSL signal loss is smaller and reaches the background level at approximately 220 °C. These results correspond well to the thermoluminescent glow curve presented in the inset graph of Figure 6. The first rapid decrease is probably related to TL peak at 157 °C and the second part of thermal signal loss can be correlated with the second TL peak (211 °C). Presented data indicates that temperature equal to 250 °C is enough to fully anneal the LSO:Ce samples.

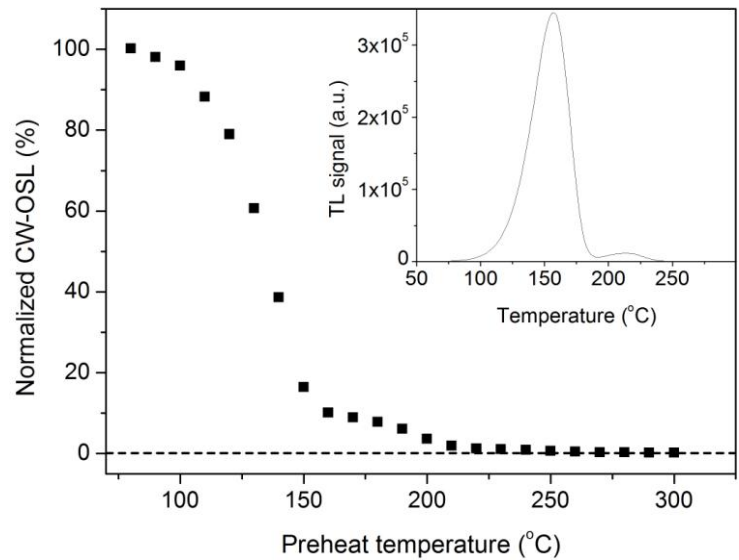

Fig. 6. Thermal stability of CW-OSL signal. Dashed line shows the background level. The inset graph present thermoluminescent glow curve of LSO:Ce sample.

The effects revealed by the preheat study entail also the fading properties of the LSO:Ce samples. The OSL signal correlation with quite low temperature TL peak results in high level of fading. Figure 7 presents the data obtained from the fading studies. Presented results were normalized to the signal received immediately after exposure to radiation. As can be seen the highest loss of the signal occurs after the first hour and it does not exceed 10%. After that time, the decrease of CW-OSL integral is slower, in particular 85% of begin signal remains after 18 hours. In all previously described experiments OSL readouts were conducted immediately after exposure. In a case of fading the significant period of time between irradiation and readout indicated that the contribution of a dose from the radioactive isotope of lutetium in the measured signal may be not negligible (Hazelton et al. 2010). In natural lutetium the abundance of beta emitting isotope Lu-176 is 2.59%. The signal of a non-irradiated sample measured after 24h was found not to exceed the background level, so the correction for self-irradiation was not necessary.

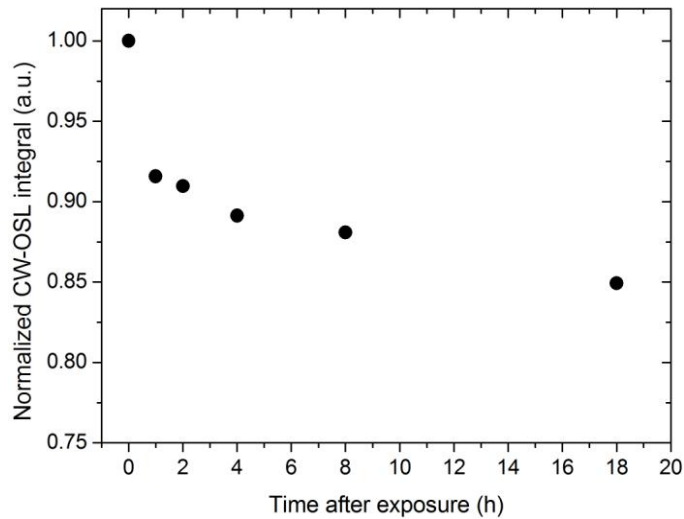

Fig. 7. Fading of CW-OSL signal. Results were normalized to signal received immediately after exposure to radiation.

**Conclusions**

It was shown that LSO:Ce possesses significant OSL signal under both green and blue light stimulation. The initial signal intensity is higher than that of commonly used OSL material $Al_2O_3$:C, but due to faster decay, the difference in signal integrals is smaller. The LM-OSL glow curve under blue light stimulation can be fitted with three first order kinetics peaks: fast, medium and slow. The dose response characteristic is linear in wide range (100 µGy - 1 Gy). The doses below 100 µGy can be easily measured. Bleaching using blue light erases all OSL signal only after 5400 s of stimulation. The temperature about 250 °C is enough for annealing process. The signal depletion during preheat is correlated with TL glow curves. CW-OSL integral is decreasing after raising the temperature to 100 °C. This results in quite high fading which is 15% after 18h.

**Acknowledgements**

This work was supported by the National Science Centre (project number 2012/05/N/ST8/03334) and NATO (project CBP.NUKR. CLG984305)